\documentclass[12pt,preprint]{aastex}
\shorttitle{Frequent Occurrence of High-speed Mass Downflows}
\shortauthors{Shimizu et al.}

\begin{document}


\title{Frequent Occurrence of High-speed Local Mass Downflows 
on the Solar Surface}


%\author{T. Shimizu}
%\affil{Institute of Space and Astronautical Science,
% Japan Aerospace Exploration Agency, 
% 3-1-1 Yoshinodai, Sagamihara, Kanagawa 229-8510, Japan}
%
%\email{shimizu.toshifumi@isas.jaxa.jp}
%
%\author{B. W. Lites, M. Kubo}
%\affil{High Altitude Observatory, National Center for Atmospheric Research,
%  P.O.Box 3000, Boulder, CO 80307, USA}
%
%\author{Y. Katsukawa, K. Ichimoto, Y. Suematsu, S. Tsuneta}
%\affil{National Astronomical Observatory of Japan, 
%  Mitaka, Tokyo 181-8588, Japan}
%
%\author{S. Nagata}
%\affil{Kwasan and Hida Observatories, Kyoto University,
% Kamitakara-cho, Takayama, Gifu 506-1314, Japan}
%
%\and
%
%\author{R. A. Shine, T. D. Tarbell}
%\affil{Lockheed Martin Solar and Astrophysics Laboratory,
% Bldg. 252, 3251 Hanover St., Palo Alto, CA 94304, USA}

\author{T. Shimizu\altaffilmark{1}, B. W. Lites\altaffilmark{2}, 
  Y. Katsukawa\altaffilmark{3}, K. Ichimoto\altaffilmark{3}, 
  Y. Suematsu\altaffilmark{3}, S. Tsuneta\altaffilmark{3},
  S. Nagata\altaffilmark{4}, M. Kubo\altaffilmark{2},
  R. A. Shine\altaffilmark{5}, and T. D. Tarbell\altaffilmark{5}}
\altaffiltext{1}{Institute of Space and Astronautical Science,
 Japan Aerospace Exploration Agency, 
 3-1-1 Yoshinodai, Sagamihara, Kanagawa 229-8510, Japan.
 shimizu.toshifumi@isas.jaxa.jp}
\altaffiltext{2}{High Altitude Observatory, 
  National Center for Atmospheric Research,
  P.O.Box 3000, Boulder, CO 80307, USA.}
\altaffiltext{3}{National Astronomical Observatory of Japan, 
  Mitaka, Tokyo 181-8588, Japan.}
\altaffiltext{4}{Kwasan and Hida Observatories, Kyoto University,
 Kamitakara-cho, Takayama, Gifu 506-1314, Japan.}
\altaffiltext{5}{Lockheed Martin Solar and Astrophysics Laboratory,
 Bldg. 252, 3251 Hanover St., Palo Alto, CA 94304, USA.}


%\altaffiltext{1}{Visiting Astronomer, Cerro Tololo Inter-American Observatory.
%CTIO is operated by AURA, Inc.\ under contract to the National Science
%\altaffiltext{2}{Society of Fellows, Harvard University.}
%\altaffiltext{3}{present address: Center for Astrophysics,
%    60 Garden Street, Cambridge, MA 02138}
%\altaffiltext{4}{Visiting Programmer, Space Telescope Science Institute}
%\altaffiltext{5}{Patron, Alonso's Bar and Grill}


\begin{abstract}

We report on  new spectro-polarimetric measurements with simultaneous
filter imaging observation, revealing the
frequent appearance of polarization signals indicating 
high-speed, probably supersonic, downflows that are associated 
with at least three different configurations of magnetic fields 
in the solar photosphere. The observations were carried out with
the Solar Optical Telescope onboard the {\em Hinode} satellite.
High speed downflows are excited when a
moving magnetic feature is newly formed near the penumbral boundary 
of sunspots. Also, a new type of downflows is identified at the edge
of sunspot umbra that lack accompanying penumbral structures. These
may be triggered by the interaction of magnetic fields sweeped 
by convection with well-concentrated magnetic flux. Another class of high speed
downflows are observed in quiet sun and sunspot moat regions. These are closely related
to the formation of small concentrated magnetic flux patches.
High speed downflows of all types are transient time-dependent mass
motions. These findings suggest that the excitation of supersonic 
mass flows are one of the key observational features of the 
dynamical evolution occurring in magnetic-field fine structures 
on the solar surface.
\end{abstract}


\keywords{Sun: atmospheric motions --- Sun: magnetic fields --- 
  Sun: photosphere --- sunspots}



\section{Introduction}

Recent high-spatial resolution imaging observations of the sun from 
ground-based facilities using advanced techniques, such as adaptive 
optics and image processings, have been revealing that the solar 
photosphere is highly structured with many large and small 
scale mass motions.
%% suggestion - this sentence probably not needed
Granules or granular cells are the most visible manifestations
of small scale mass motions on the photosphere and are formed by
convection.
Magnetic fields of the sun, some of which are observed as 
tiny bright points in G-band images \citep{ber96}, wander
due to interaction with 
nearby evolving granular cells \citep{ste98}. The interaction 
between granular motions and magnetic field can provide kinetic 
energy to the magnetic field and drives much more dynamic behaviors 
as well as plasma heating in the solar upper atmosphere. Sunspots,
the largest visible magnetic structures at the photosphere, 
also show Evershed mass flows in the radial direction of the penumbra 
\citep{sol03}. It is not easy with ground-based observations 
to trace dynamical behaviors of mass motions and their association 
with evolution of magnetic fields and heating 
in the upper atmosphere at sub-arcsecond 
spatial scales because of variable seeing conditions. Therefore we
lack observational knowledge about the overall dynamical 
evolution of the solar atmosphere including mass motions, magnetic 
fields, and heating. It is apparent that our observational
knowledge can be much improved via high spatial-resolution 
observations from space with seeing-free stable conditions. 
New observations of dynamic flows associated with magnetic fields
would provide key clues toward understanding the nature of
atmospheric dynamics and thus the physical mechanisms
that control the dynamical evolution of magnetic field features. 

Spectro-polarimetric observations of magnetically sensitive spectral lines in
visible and infrared wavelengths have become crucial in the past decade 
for quantitative investigations of magnetic field structures in 
the photosphere \citep{lit93}. 
The spectro-polarimetric observations
also probe the line-of-sight Doppler motion of 
the mass in the atmosphere occupied by the magnetic field.
The Solar Optical Telescope (SOT) \citep{tsu07, tar07, ich07, sue07, shi07a} 
onboard a new Japanese spacecraft {\em Hinode} \citep{kos07}, 
starting its scientific 
observations from November 2006 after its successful launch on 22 
September 2006, includes the Spectro-Polarimeter (SP) instrument, which
performs for the first time precise (0.1\%) measurements of 
polarization with high spatial resolution (0.3 arcsec) under seeing-free 
conditions. The SP observations provide highly dispersed spectral profiles 
of the polarization (Stokes profiles) 
suitable for investigating mass flows associated with magnetic field. 

One of the remarkable features observed in the Stokes profiles 
is the frequent appearance of complex Stokes V (circular 
polarization) profiles showing signal excess at the red side (longer wavelength) 
of the Fe I 630 nm spectral lines, as shown in Figure 1. 
At two local positions, a signal elongating toward 
the right is clearly seen $250-400$ m\AA\ from the center 
of the spectral lines. 
Since the peak of the antisymmetric Stokes V profile is roughly located 
100 m\AA\ from the line center, a $250-400$ m\AA\ offset corresponds 
to 7-14 km s$^{-1}$, suggesting mass
flows exceeding the sound velocity in the photosphere (6-7 km s$^{-1}$)
associated with the magnetic field. Based on Stokes profile measurements,
the number of observational reports on supersonic or rapid flows of 
any type in the photosphere is recently increasing;
a strong downflow close to the neutral line of a delta sunspot 
\citep{mar94, lit02}, supersonic Evershed downflows in a sunspot 
\citep{wie95, del01},
strong downflows at the location of chromospheric Ellerman bombs \citep{soc06}, 
magnetic elements amplified by flux expulsion and convective collapse 
\citep{sig99, bel01, soc05}, magnetic elements in
quiet and active regions \citep{sig01}, supersonic downflows in 
the sunspot moat \citep{shi07b} and so on. It appears from these 
previous observations that supersonic mass flows occur in localized
areas, which can be revealed only with high spatial resolution 
observations. The SOT
observations can provide time series of spectro-polarimetric 
measurements with sub-arcsec spatial resolution under stable seeing-free
condition, allowing us to investigate 
the dynamical evolution of high speed flows and its relation to
features seen in other imaging observations.

This paper presents new {\em Hinode} observations showing that supersonic 
downflows frequently occur at various magnetic circumstances 
in the photosphere (section 3). In section 4, a few typical examples
corresponding to each of downflow events classified in section 3 are 
shown to trace temporal evolution of the events and their associated
features captured with simultaneous filter imaging observations, followed
by discussions in section 5.
These observations show the importance of further investigations on 
the supersonic downflows for understanding fine structures and their dynamic 
evolution of magnetic field concentrations as well as for understanding 
transient heating at the chromosphere.

\section{Observations}

Two types of spectro-polarimetric observations were made with the
Spectro-Polarimeter of SOT on December 2, 2006. One was a fast mapping
observation performed between 0:00-3:30 UT and the other is a
normal mapping observation performed between 10:30-15:30 UT. 
The primary purpose of this fast mapping observation was to capture overall
temporal signatures of rapid flows in the active and quiet 
region with a field of view as large as possible.
The fast mapping mode has a 4.4 minute cadence 
for a 20.5 arcsec-wide scan with 0.32 arcsec
effective pixel size (64 positions in east-west direction). 
In the fast mapping, since the slit width is 0.16 arcsec, photons 
are accumulated during one modulator rotation (1.6 sec) at 
the 1st slit position and during another rotation at the next slit position.
The field of view along the slit is 82 arcsec (256 pixels). 
The supreme spatial resolution capability of the SOT is not
fully exploited with the fast mapping observation, and instead 
the normal mapping observation provides the best spatial 
resolution (0.16 arcsec pixel) 
data. This particular normal 
map mode had a 67 sec cadence for a 2.1 arcsec-wide scan
(13 positions, 4.8 sec photon accumulation for each) in 
east-west direction with 41 arcsec (256 pixels) along the slit.
Hence it covered a much smaller area than the fast maps but
with higher resolution and longer integrations.

A moderate sunspot was partially included in the field of view 
with a moat flow area and quiet region located north of the spot (Figure 2).
The spot is a leading spot of active region NOAA 10926, 
located near the disk center (200 arcsec S, 200 arcsec W), and it 
consists of three umbrae divided by light bridge structures.

The SP obtains the full polarization states of line profiles of two 
magnetically sensitive Fe lines at 630.15 and 630.25 nm and nearby 
continuum. 
We define a measure to identify the elongated signal excess located far from 
the line center by integrating the Stokes V signals
over $259$ to $431$ m\AA\ from the center. 
The zero-velocity is derived from the profile 
averaged over all the Stokes I profiles along the slit 
(256 profiles).
This measure has the advantage of detecting the complicated Stokes V profiles
that suggest the existence of high speed signals. Note that
the zero-crossing wavelengths of Stokes V profile \citep[e.g.,][]{sig01}
sometimes used for quantitative studies do not detect high speed components 
adequately for some types of Stokes V profiles, if a largely shifted
component has the same polarization polarity as the rest component
(See section 4 for examples).

During the normal-mapping spectro-polarimetric measurements, 
the Broadband Filter Imager (BFI) of SOT produced photometric images
in the G (430.50 nm center, 1.2 nm width) and Ca {\sc II H} (396.85 nm
center, 0.3 nm width) spectral bands at the highest spatial resolution
(about 0.2 arcsec, 0.0545 arcsec in pixel size) every minute.

\section{Frequent Occurrence of High-Speed Downflows}

Figure 2 shows a time series of our measure
representing signal excess for the red and blue sides 
of the spectral line.
A large number of point-like, 1 arcsec or less in diameter, dark
and bright signals (opposite signs of V) can be easily identified in the red-side wing
at various locations.  % in the red-side maps.
The point-like signals have durations of 4 - 30 minutes, most 
only about 4-10 minutes, indicating that all these high-speed
downflows in the photosphere are transient and not static.

Three types of high-speed downflows can be identified, each
associated with their spatial relationship to the sunspot: 
1) Around the outer boundary 
of sunspot penumbra. Most of these events are located in a moat flow
region surrounding the penumbra, and some events are also seen
from the middle to the outer edge of penumbral structures \citep{del01}. 
Dark features are slightly predominant, 
i.e., negative integrated Stokes V
at the red side of the spectral lines.
2) At the edge of those portions 
of umbrae that do not have associated penumbral structures. 
These transiently appear with a strong bright signal in the map. 
In the sunspot examined in this paper, the portions of umbrae
without penumbral structures are formed in association with
light bridge structures where evolving granular cells 
are present. 
Another portion of the umbra without associated penumbral 
structures is seen at the lower right-hand-side corner of 
the field of view where the
umbra is face to face with a different sunspot with the same
magnetic polarity located to the south. 
3) At small magnetic field
concentrations located in moat and quiet regions far from the 
sunspots. In many events, a bright point with sub arcsec diameter 
is observed in the Stokes I continuum and more clearly in G-band images 
at the exact location of downflows.

It should be pointed out that strong signal excesses are not
as common in the blue side of the spectral lines, except for a few
events at the small magnetic field concentrations located near the outer
boundary of penumbra. This indicates less upward mass motions exceeding 
the sound velocity at the photospheric level.

Figure 3 shows the full Stokes profiles typically observed for 
the three types.
At the outer edge of penumbral (type 1),
a common characteristic of the Stokes V profile is
a complex shape with three lobes, unlike the usual, antisymmetric
profile with two lobes. 
The two lobes located at both sides about equidistant
from the zero velocity wavelength have
a symmetric Stokes Q/U (linear polarization) shape
with a signal strength similar to that of Stokes V. 
This indicates an inclined magnetic field at rest.
The extra lobe is observed at the red side with 
predominantly negative signal and with no clear
corresponding signal in Stokes Q and U, suggesting
that the magnetic field forming the third lobe is primarily
vertical.

The second type, observed at the umbral edge without accompanying penumbra, 
also shows a complex Stokes V profile with three 
lobes (Figure 3b).
The unique signature common to these events
is that the third lobe is positive, 
indicating a magnetic field
aligned to the ambient magnetic field of the sunspot.
This type of motion is also observed around the outer boundary 
of pores, which also, of course, do not have associated penumbral structures.

The third type is observed in the moat and quiet area 
where small concentrations
of magnetic field are scattered.
Some of the Stokes V profiles have an elongated signal toward
the longer wavelength in addition to the usual, antisymmetric
profile with two lobes (Figure 3c). In some cases there is 
an extreme red-shifted lobe, similar to that of the umbra.
No linear polarization is observed, indicating that the
magnetic field is vertical.

Here we note a limitation of the measure used in 
this analysis. Small values of this quantity do not necessarily
imply the absence of high-speed downflows. In fact, weak Stokes V signals 
will always produce small values of this quantity even if strong 
downflows exist. On the other hand, strong Stokes V signals can 
produce large values of this quantity even if a strong downflow exists 
only in a small fraction of the pixel. Thus, the magnitude of 
the measure should be interpreted carefully, 
although the measure is useful to easily identify 
high-speed downflows that are associated with magnetic flux.

\section{Temporal Evolution and Associated Features}

In this section, we examine the temporal development
of high-speed downflows using the Stokes profiles 
and simultaneous filter imaging observations to understand 
how high-speed downflows develop with their associated 
magnetic fields.

\subsection{High-speed downflows near sunspot penumbral boundaries}

Figure 4 shows the temporal evolution of typical downflow events
observed just outside the sunspot penumbral boundary. 
The field of view of each frame is $3.8$ (N-S) $\times 3.3$ (E-W) arcsec. 
Note that the quantities derived from the normal-mapping Stokes measurement
are 2.1 arcsec wide in the east-west direction.
High-speed downflows (negative signal) are observed twice in this 
sequence (16 min coverage), each of
which is accompanied by a tiny positive signal. 
The second event is marked by circles with a radius of 1.0 arcsec
in Figure 4. They are located
almost at a short extension (1-2 arcsec) from a dark filamentary structure 
of the penumbra, where nearly horizontal Evershed gas flows seem
to be preferentially concentrated with nearly horizontal magnetic 
field \citep{tit93, shi94, rim95, rou02, bel06, ich07b}. Note that
the tip of the dark filamentary structure is seen at the bottom of
Stokes I continuum and G-band images, as identified with the contour lines
in G-band. The duration of the events is 3-4 minutes. The longitudinal
magnetograms, generated by integrating the blue wing of Stokes V, 
show that a white patch newly appears at almost the same time (t= 539s) 
and location of the second event. The white patch at t= 539, 607, and 
674s is still negative-polarity, but it can be recognized as a white 
patch because this negative polarity signal is much weaker than 
the surrounding negative-polarity signals. 
The center portion becomes truely positive polarity at t= 741s and 
the positive polarity signal continues to develop after the time,
finally forming a smaller concentrated positive patch. 
This evolution of the white patch suggests 
that a positive-polarity magnetic flux starts to develop in 
the negative-polarity flux dominant area from around t = 539s, which is 
when the second high-speed downflow event begins.

The positive patch slowly moves outward (upward in the figure), 
approaching a pre-existing negative-polarity patch, where tiny positive 
high-speed signals are observed at the red wing of the Stokes V profiles. 
The magnetic flux in the negative-polarity patch seems to slowly 
decrease with time. The positive-negative pair of the magnetic features
can be recognized as a bipolar moving magnetic feature 
\citep[MMF, e.g.,][]{har73}. Note that the temporal evolution shown here
is rather common in the formation of MMFs and another example can 
be found in \citet{kub07a}. The positive magnetic patch with strong
downflow exists in an inter-granular lane at the front edge of 
a granule which gradually expands outward from the penumbral side. 
The series of G-band images show that a bright compact feature starts 
to appear and brighten from t= 270s and another bright feature from
t= 539s near the tip of the dark 
filamentary structure and it seems to be pushed by an expanding granule. 
Moreover, Ca {\sc II H} images show a transient brightening
at the location of downflow events. 
As clearly seen in the second event marked
with circles, a small patchy bright feature appears at the same time
as the downflow event (t= 539s) and fades after t= 876s. 
This signature is clearly different from longer-lived enhanced 
Ca {\sc II H} emission seen in strong flux concentrations.

This example clearly shows that
high-speed downflows are excited at the extension from
Evershed flow channels in the penumbra, resulting in a vertically
oriented magnetic field, well known as an MMF. Simultaneously, transient 
heating occurs in the chromospheric layer, which may mean that
the magnetic field of the newly appeared positive-polarity field expanded and
reached the chromosphere. Also, it might be indirect
evidence of heating due to shocks formed by supersonic downflows.

Figure 5 shows the temporal evolution of full Stokes profiles at the location
of the second large downflow event. The Stokes profiles correspond to 
the center of the downflowing structure defined by the center of the 1.0-arcsec-radius 
circles in Figure 4. For the profiles before and after the downflow 
event, we traced the magnetic signature observed in G-band and magnetic 
flux images. Before t= 270s, the position is unchanged from the last 
identified position, because it is difficult to trace the magnetic 
signature. The third lobe starts to appear in Stokes V from
t= 472s and is well observed until t= 741s. The profile of 
Fe {\sc I} 630.15nm at t= 809s still shows a conspicuous third lobe.
The high-speed downflow forming
the third lobe reaches 9.2 km s$^{-1}$ at t= 674s, derived from 
the wavelength of a zero-crossing formed with the third lobe. 
The red and blue wing amplitude of antisymmetric Stokes V
profile decreases during the period when the third lobe is observed
and a Stokes V signal with the opposite polarity amplitude appears by
the end of the sequence, 
indicating that the line-of-sight magnetic polarity of the main
magnetic field component is completely replaced with the opposite polarity
magnetic field formed with the excitation of high-speed downflow. 
The linear polarization (Stokes Q and U) signals shows that the inclined
magnetic field evolves to almost vertically oriented field
when the high-speed downflow is well observed (t= 607 - 741s). 
An inversion based on the Milne-Eddington model approximation gives
an inclination of the magnetic field of 114 deg with strength
of 809 gauss at t= 0s, where the field inclination of 90 deg is
defined as horizontal. At t= 944s, the magnetic field evolves to 
1975 gauss with an inclination of 50 deg.

\subsection{High-speed downflows at sunspot umbral edge}

Figure 6 shows the temporal evolution of typical downflow events
observed at the sunspot umbral boundary. Note that the events are
located near the lower edge of the Stokes observing region.
Three high-speed downflow events are detected in this area; one
bright event is observed  in the upper left of the
frames (just outside the circles) over t= 270 - 472s, 
the second event is at the left side of the circles over t= 203 - 809s, and
the third bright event starts to brighten from around t= 472s,
and is marked by circles with a radius of 1.0 arcsec.
These high-speed 
downflow events are located at the boundary between the dark 
umbral region and bright granules, which can be identified 
by the contours. 
Radial outflows, the well-known sunspot moat flows, are not seen
at umbral edges without penumbra, as shown in \citet{var07}.
The size of nearby granules appearing beside
the umbral edge is much smaller than those seen in the quiet Sun,
which is reflected by the distribution of strong magnetic flux
not only inside the umbral region but also in the nearby granules.
The magnetic flux map shows that strong flux of the umbral region
is extended to the nearby granule region.
The nearby granules evolve during the time that the high-speed downflows 
are observed, although no clear correlation between the granule evolution
and downflow development is derived from this observation. 
No clear features are
observed in the series of Ca {\sc II H} images. 

Figure 7 is the temporal evolution of the full Stokes profiles at the location
of the third downflow event. The 
profiles are for the center of the 1.0-arcsec-radius 
circles in Figure 6. Before the downflow event, 
the position is unchanged from the position identified at t= 472s, 
because it is impossible to trace the magnetic signature. 
The antisymmetric Stokes V profile with two lobes does not show 
significant changes during the period, but an elongated signal 
toward the longer wavelength appears from t= 405s. The Fe {\sc I}
630.15 nm line shows that the elongated signal can be identified as 
the third lobe in t= 539 to 741s frames. 
The scale of the vertical axis in Stokes V
is 4 times larger than in Stokes Q and U, meaning that the magnetic
field has almost a vertical orientation. 
The Milne-Eddington inversion for the profiles at t= 944s gives
a 156 deg inclination for a 1685 gauss
magnetic field.

\subsection{High-speed downflows at moat and quiet regions}

Figure 8 shows the temporal evolution of one typical downflow event
observed in the quiet region far from the sunspot. The downflow signal
is observed in a weakly concentrated negative-polarity patch, which
is located in an inter-granular lane. It seems that the negative-polarity
patch is rather diffuse, about 1 arcsec in size at t = 203 s,
and it developes to form a concentrated magnetic flux less than
0.5 arcsec in size. When the concentrated magnetic flux 
appears in longitudinal magnetogram (after t= 337s), the signal excess
at the red wing is well observed, followed by the disappearance of
the signal excess (t= 876s). The series of G-band images show that
a bright point \citep[GBP, e.g.,][]{ber96} is newly 
formed after t= 337 s at the exact location of
the signal excess and magnetic flux concentration. The size of GBPs 
is 0.3 arcsec in diameter. Immediately after starting 
the excitation of a high-speed downflow, a point-like transient brightening is 
observed in Ca {\sc II H} images (after t= 405s).
After the downflow signal disappears, the chromospheric brightening 
signature in Ca {\sc II H} also fades out (t= 944s).

The temporal evolution of full Stokes profiles at the
centers of the 1.0-arcsec-radius circles
are given in Figure 9.  
No apparent signal is seen in Stokes Q and U above the noise level, 
meaning that the magnetic field is vertically
oriented. The third lobe is well seen for more than 10 minutes from the 2nd 
profile (t= 67s), especially in the Fe {\sc I} 630.15 nm line, whose line core is
formed slightly higher than that of 
Fe {\sc I} 630.25 nm line in the solar atmosphere. 
Since the third lobe has the same polarity as the red wing lobe of
the usual, antisymmetric Stokes V profile, the third lobe does not make
the second zero-crossing. But the Stokes V profile of Fe {\sc I} 630.15 nm
line at t= 539s shows a zero-crossing by the third lobe, giving
7.8 km s$^{-1}$ (164m\AA\ from the line center), which is supersonic.
The blue wing amplitude of the antisymmetric Stokes V profiles 
shows a gradual growth
as a function of time, suggesting the increase of either (or both) magnetic 
field strength or magnetic filling factor (how much area is 
occupied by the magnetic field). Actually, the Milne-Eddington inversion
gives that the intrinsic magnetic field strength changes from 479 gauss
(t= 0s) to 1514 gauss (t= 944s) and the magnetic filling factor from
0.22 (t= 0s) to 0.35 (t= 944s).

\section{Discussions}

\subsection{High-speed downflows associated with formation of MMFs}

It was shown that the excitation of high-speed downflows is 
observed outside the sunspot penumbra when a magnetic flux patch with 
a polarity opposite to the sunspot is newly formed. Such
patches may be identified as one polarity patch of a MMF. MMFs sometimes 
have bipolar magnetic structures and stream away from near the outer 
boundary of the penumbra of well-developed sunspots  
\citep{zha03, hag05, cab06, kub07b}.
This observation suggests that bipolar MMFs
may be effectively developed and formed in association with 
the excitation of high-speed downflows in a magnetic tube.
An outward directed flow, Evershed flow, is well observed in 
the sunspot penumbra, and downflows are observed at some locations 
at the outer edge of the penumbra, returning the mass flux 
below the photosphere \citep{wes97, sch02, bel04, bor06, bec08}. 
The Evershed flows along individual channels are transient with 
a timescale of the order of 10-15 minutes \citep{shi94, rim94}, and
some of the downflows can evolve to supersonic flows at 
the photospheric layer \citep{del01, bel04}.
The observed high-speed downflows
are predominantly located around the outer edge of the penumbra and its
extension to the moat flow area. Closely related to downward magnetic 
pumping \citep{tho02}, pairs of magnetic features may be more effectively 
developed by downward kinks formed by a high-speed downflow.

Before the excitation of a high-speed downflow, there is an inclined
magnetic field in the moat region, which is partially an extension 
from 
nearly horizontal penumbral structures of sunspots. Transient Evershed flow gas
along the penumbral horizontal field may continue to flow along 
the inclined magnetic field located just outside the penumbral
boundary. A bulk of gas supplied to a small height 
above the photosphere by a transient Evershed flow may start to drop into
the photosphere along magnetic fields due to gravitational force. 
Such a bulk flow may reform a part of magnetic 
field to a more vertical oriented configuration and 
the gas may undergo a free fall,
resulting in accelerating to high speed.
At this time, a complicated Stokes V 
profile with three lobes is observed; The observed spectral profiles
can be interpreted with two magnetic components existing within 
0.3 arcsec resolution or with two magnetic components existing
along the line of sight in the thin layer of the line
formations, as shown in Figure 10a. 
The main component is a pre-existing inclined magnetic field with 
no strong mass motions, which represents the global structure of the
magnetic field at the outer area of penumbra. Moreover, because of the
highly red-shifted nature of the third lobe, a limited 
portion of magnetic field carries a high-speed downflow 
forming a vertical field orientation. The downward mass motion 
continuing for roughly 5 minutes reforms the magnetic field 
into more vertically-oriented magnetic field in the opposite direction of
the spot. 
After ending the high-speed downflow, the opposite-directed 
vertically-reformed magnetic field may go on to form a strong
concentrated flux, as described in section 5.3.

\subsection{High-speed downflows associated with convection in strong magnetic fields}

High-speed downflows are sometimes observed at the edge of sunspot
umbra without accompanying penumbral structures. Also, similar downflows
are observed at the edge of pores, which are small concentration of
magnetic flux without horizontally oriented penumbral structures.
An expanding granular cell with significant magnetic flux is located beside the umbra.
High-speed downflows are observed at the interface between the umbra and the granular cell.
A significant circular polarization signal is emitted from 
the granular cell, indicating vertically oriented magnetic flux. 
The development of a granular cell may force
magnetic fields to sweep toward the umbral boundary, which may
trigger the interaction of the sweeping field with well-concentrated 
magnetic flux in the umbra (Figure 10b). One possible interaction is 
a magnetic reconnection between them. 
To form a discontinuity between the magnetic fields, the sweeping field
needs to be inclined to the magnetic field inside the umbra.
As a result of magnetic reconnection
in the photosphere or lower chromosphere, a reconnection outflow directed
toward the surface may be observed as supersonic mass motions. 
However, it is a puzzle why strong upward motions are not observed.
The other possible interaction is that the physical process described in
section 5.3 are excited when the sweeping magnetic field is merged to
the pre-existing umbral field with kilo gauss strength.

\subsection{High-speed downflows at quiet and moat regions}

The example shown in Figure 8 and 9 clearly shows that high-speed 
downflows are excited in the vertically-oriented magnetic field area
during the formation of a small concentrated magnetic 
flux patch, which can also be seen as a bright point in G-band. 
This observation is consistent with previous observations 
reporting on supersonic downflows at the magnetic elements amplified by flux 
expulsion and convective collapse \citep{sig99, bel01, soc05}.
Convective collapse is theoretically introduced as a mechanism for 
concentrating magnetic fields into thin ($\approx 100$km, or 0.1-0.2 arcsec 
in diameter) but strong (kilo gauss) flux tubes \citep{par78, spr79}. 
It is predicted in the theory that a downward flow can be excited due 
to convective instability developed by radiative cooling inside 
the flux tube with decreasing heat input from ambient atmosphere 
across magnetic field lines. The high-speed downflow leads to a partial
evacuation of the gas and therefore a concentration of the magnetic
flux with increasing magnetic field strength (Figure 10c). 
This paper is mainly focused on the mass motions inside the concentrated
magnetic flux patch observed in full Stokes profiles. 
Careful multicomponent inversions of these Stokes 
spectra will be required for precise determinations of
the physical properties inside the flux tubes with high speed mass flows. 
It should be noted that another example from a {\em Hinode}
spectro-polarimetric observation is recently examined with an 
inversion technique \citep{nag07}.

The observed high-speed downflows may be more effectively initiated if the gas
is drained from the low chromosphere into the photosphere. We have observed
flux emerging all the time in quiet areas and plage regions \citep{cen07, ish08}.
The emerging flux will carry mass upward, which must flow downward 
at some point. The upward motions of the emerging magnetic fields may also 
constrict magnetic flux tubes so that the gas flows could become high 
speed in the tubes and the high-speed flows are sustained in the photosphere.
Indeed, it is observed that the Stokes V profile is strongly red-shifted 
at one end of the emerging, horizontal magnetic structure with velocity 
up to 5 km s$^{-1}$ \citep{ish08}.

Almost simultaneously with the excitation of a high-speed downflow, 
a point-like transient brightening is also observed at the exact location 
of the high-speed downflow with the Ca {\sc II H} filter.
The Ca {\sc II H} filter has a response function 
with an average response height of 247 km above the photospheric level 
and a long tail extending into the middle chromosphere \citep{car07}. 
Transient Ca {\sc II H} brightenings can be interpreted by various kinds 
of mechanisms including mechanisms discussed below. 
One interpretation is that the Ca {\sc II H} brightening is a signature of
transient heating of the chromospheric gas either by a heat source located 
in the chromosphere or energy transfer from the photosphere through 
magnetic fields. Another interpretation is that the observed Ca {\sc II H} 
brightening is due to decreasing the line absorption with reduced opacity 
of the gas. This may be caused by a strong evacuation inside the magnetic
flux tube at chromospheric layer, if the high-speed downflows observed
at the photosphere extends to the chromosphere during 
the formation of the kG flux tubes.

The first interpretation, transient heating in the upper atmospheric 
layer, is also suggested
by an observation of red-shifted events observed at the outer moat 
region using full Stokes measurements of infrared photospheric 
Si I 10827 \AA\ line with Tenerife Infrared Spectro-Polarimeter 
\citep{shi07b}. The physical conditions of the red-shifted component
derived with SIR (Stokes Inversion based on Response functions) inversion
show that the downflowing component has a relative downward motion as
large as 10 km s$^{-1}$ with a vertical orientation of the magnetic field and
that the temperature of the downflowing component seems to be 400-500 K
higher than that of the stationary component. These observations
suggest that a magnetic reconnection occurring in the upper atmospheric
layer excites a supersonic downflow \citep{shi07b}, although
a slight delay was observed in Ca {\sc II H} brightening compared with
the excitation of a high-speed downflow.
Note that a supersonic downflow is also reported at the location of 
chromospheric Ellerman bombs \citep{soc06}, which may be a magnetic 
reconnection at the chromosphere. 

As an alternate mechanism, a shock front 
formed by a supersonic downflowing motion inside a magnetic flux tube 
may play a key role in determining the further dynamical evolution of 
the concentrating magnetic flux \citep{tak99}. According to this
simulation, the shock formation is restricted within the most 
unstable portion of the examined parameter range. A strong downflow rebounds in
the deeper layers and the resulting upward mass motion leads to
a shock wave penetrating from the photosphere to the chromosphere,
which may be observed a weak blue-shift signal in Stokes V profiles
\citep{bel01, soc05}. The energy deposition in the low chromosphere by 
the propagating shock waves may be observed as a point-like transient
brightening in Ca {\sc II H} images.

\section{Conclusions}

The Spectro-Polarimeter provides new precise observations 
of solar magnetic fields as well as velocity fields at the surface with 
0.3 arcsec spatial resolution. 
It reveals frequent occurrences of high-speed, probably 
supersonic, downflows in the photospheric layer. 
We find that the photosphere is full of extremely dynamical 
small scale structures with high-speed mass motions. 
High-speed downward mass flows occur not only in association with sunspot 
magnetic field structures but also in small concentrations of magnetic flux 
in the quiet sun. 

Understanding the sources of high-speed mass downflows needs further 
observational investigations with Stokes-polarimetric observations
by {\em Hinode}. However, the temporal evolution of
some downflow events shown in this paper suggests a few magnetic
circumstances for exciting high-speed mass flows 
at the photosphere: the extension from 
Evershed flows in sunspots, interaction between intense magnetic field
and nearby convective gas motions, and magnetic elements amplified by 
flux expulsion and convective collapse. The high-speed downflows
observed at the extension from the penumbral filamentary structures
are suggested to play an important role in forming MMFs 
around the sunspot penumbra. High speed flows in a magnetic 
flux tube may play an role in forming kinks of magnetic field.
Associated with high-speed downflows, transient brightenings are also
observed in chromospheric Ca {\sc II H} images. 
Supersonic mass motions would form shock 
fronts and generate shock waves in the atmosphere, which can be important 
in propagating the energy and the heating of plasma.

\acknowledgments

Hinode is a Japanese mission developed and launched by ISAS/JAXA, with NAOJ
as domestic partner and NASA and STFC (UK) as international partners. It is
operated by these agencies in co-operation with ESA and NSC (Norway). 
The authors would like to express their thanks to all the people who have
been involved in design, development, tests, launch operation, and science
operations for realizing the {\em Hinode} (Solar-B) mission and its new 
advanced observations presented in this paper. The authors would like to
acknowledge the anonymous referee for his comments helpful 
in improving the contents of the paper.

\clearpage

%% the file, uncomment the \plotone, \plottwo, and
%% If you need a layout that cannot be achieved with \plotone or
%% \plottwo, you can invoke the graphicx package directly with the
%% \includegraphics command or use \plotfiddle. For more information,
%% please see the tutorial on "Using Electronic Art with AASTeX" in the
%% documentation section at the AASTeX Web site,
%% http://www.journals.uchicago.edu/AAS/AASTeX.
%%
%% The examples below also include sample markup for submission of
%% supplemental electronic materials. As always, be sure to check
%% the instructions to authors for the journal you are submitting to
%% for specific submissions guidelines as they vary from

%% This example uses \plotone to include an EPS file scaled to
%% 80% of its natural size with \epsscale. Its caption
%% has been written to indicate that additional figure parts will be
%% available in the electronic journal.

\begin{figure}
\epsscale{.50}
\plotone{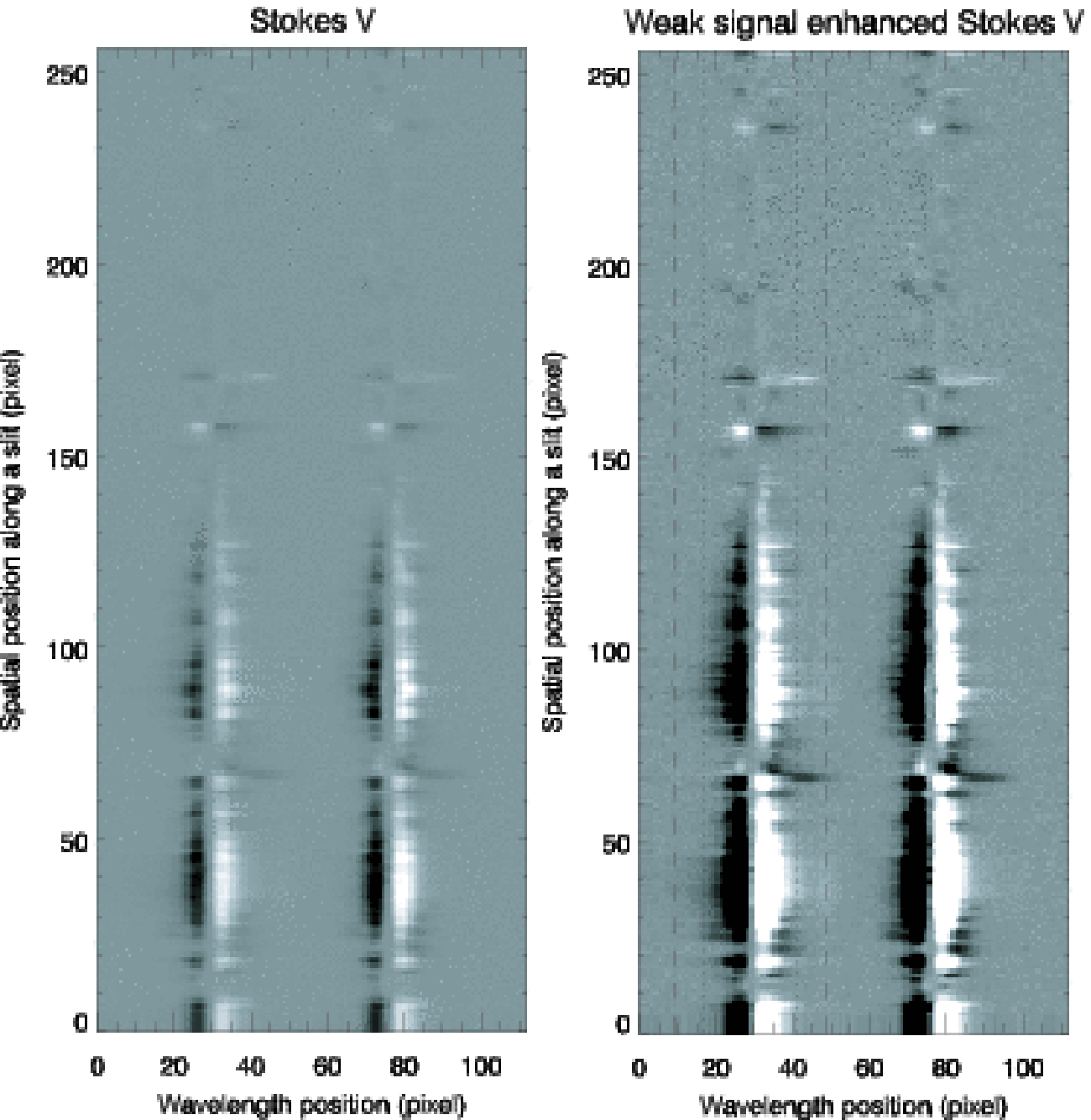}
\caption{Example of Stokes V profiles of magnetically sensitive 
    Fe I 6301.5 and 6302.5 \AA\ lines observed at a slit position.
    At the right panel, the color depth scale is changed to enhance
    weak signals seen at wavelengths far from the line core 
    (left: $-5000$ to $+5000$ DN, right: $-1000$ to $+1000$ DN). 
    Horizontal axis gives wavelength position 
    with spectral sampling of 21.549 m\AA/pixel. 
    Wavelength increases toward the right. The dotted and dashed lines 
    are 259 m\AA\ and 431 m\AA\  from the center of the averaged 
    line profile. Pixels along the slit are 0.317 arcec.}
\end{figure}

\clearpage

%% Here we use \plottwo to present two versions of the same figure,
%% one in black and white for print the other in RGB color
%% for online presentation. Note that the caption indicates
%% that a color version of the figure will be available online.
%%

%\begin{figure}
%\epsscale{.60}
%\plotone{shimizufig2.eps}
%\caption{ SP scanning area in active region NOAA 10926. The image is Stokes V
%  image acquired with Narrowband Filter Imager, another focal plane instrument
%  of the SOT, showing that dark (bright) is negative (positive) polarity 
%  of line-of-sight magnetic flux. A well-developed 
%  leading spot of the active region is captured at the lower half of 
%  the scanning area. A part of moat flow region surrounding the sunspot and 
%  quiet region extended from the moat are observed at the upper portion 
%  of the area.}
%\end{figure}

\begin{figure}
\epsscale{1}
\plotone{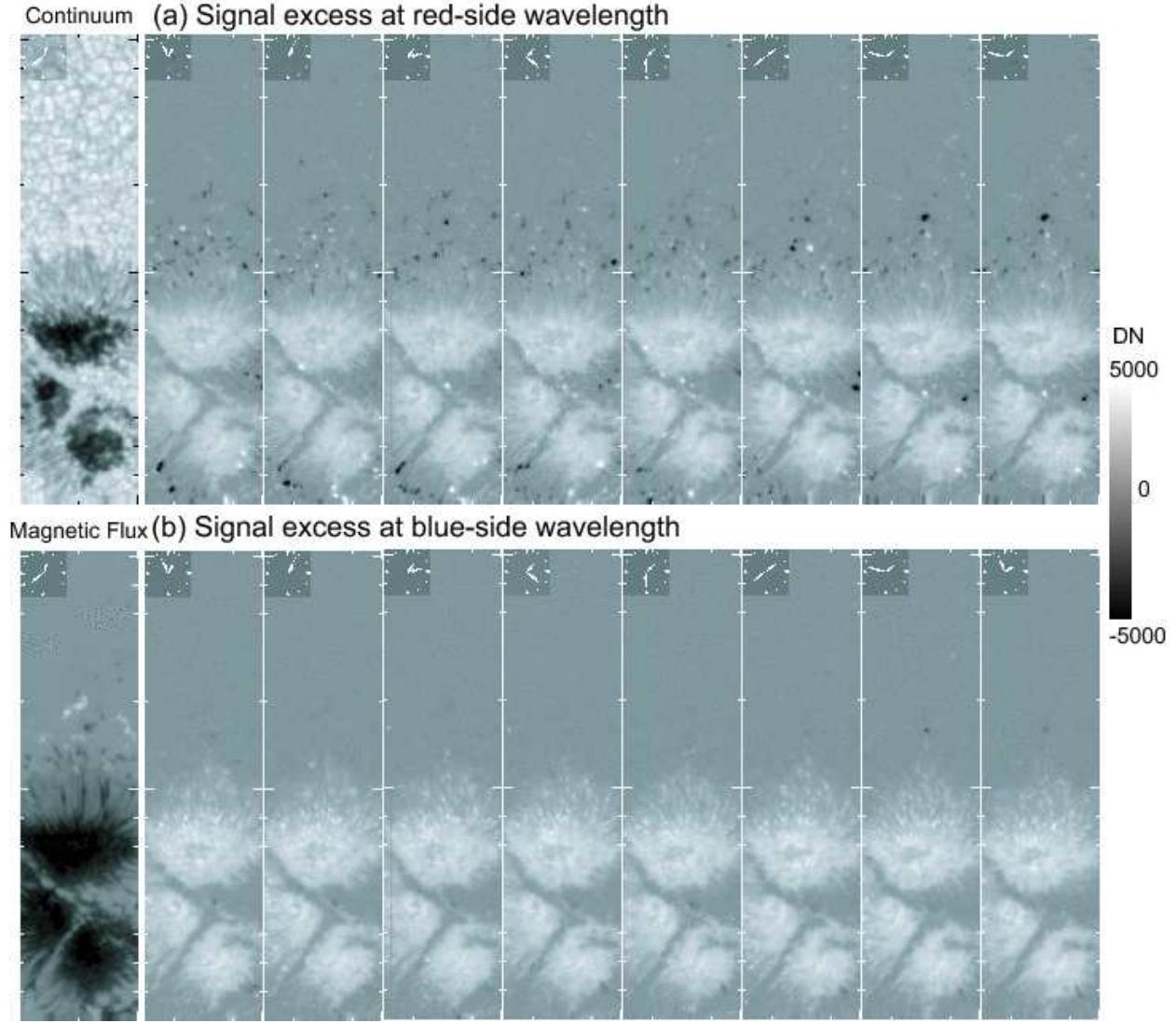}
\caption{
Time series of our measure for detecting signal excess at the red and blue sides
of the spectral line, shown with a continuum and magnetic flux image. 
The measure is the integration of Stokes V signals between $259-431$ m\AA\ from 
the center. In (a), bright features indicate positive integrated
Stokes V at the red side, whereas the polarity 
is reversed in (b) and dark features indicate positive integrated Stokes V at
the blue side. This time series of the measure is a part of the fast mapping 
observation performed from 0:00-3:30 UT on 2 December 2006. Tick marks 
are given every 5 arcsec.
%The magnetic flux image gives light-of-sight component of
%magnetic flux with while for positity and black for negative polarity. 
}
\end{figure}

\begin{figure}
\epsscale{0.7}
\plotone{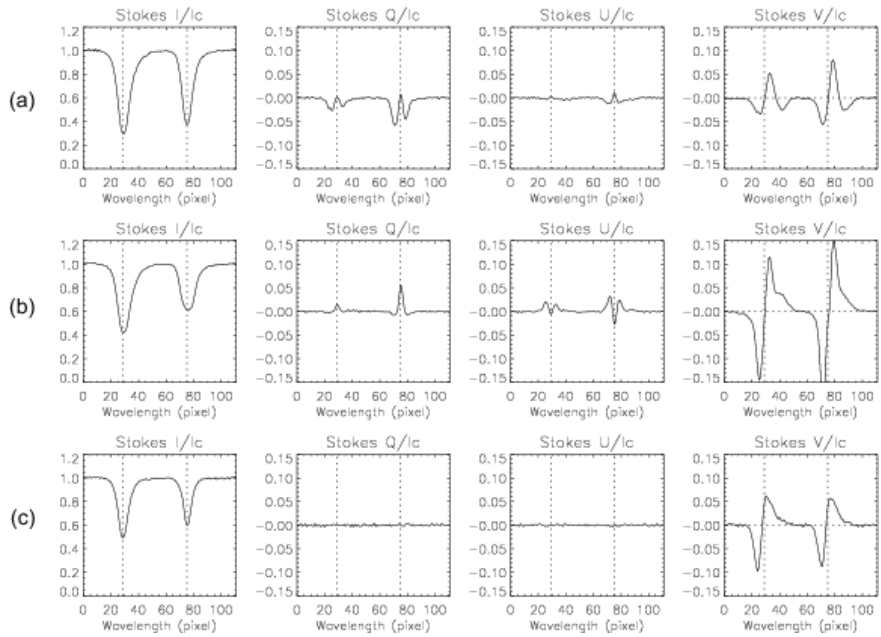}
\caption{Full Stokes profiles representing three types of high speed 
    mass downflows: (a) Near the outer boundary of
   sunspot penumbra, (b) at the umbral edge without accompanying
   penumbral structures, and (c) in magnetic patches observed
   in quiet sun. The profiles are normalized to the nearby
   continuum.}
\end{figure}

\begin{figure}
\epsscale{0.45}
\plotone{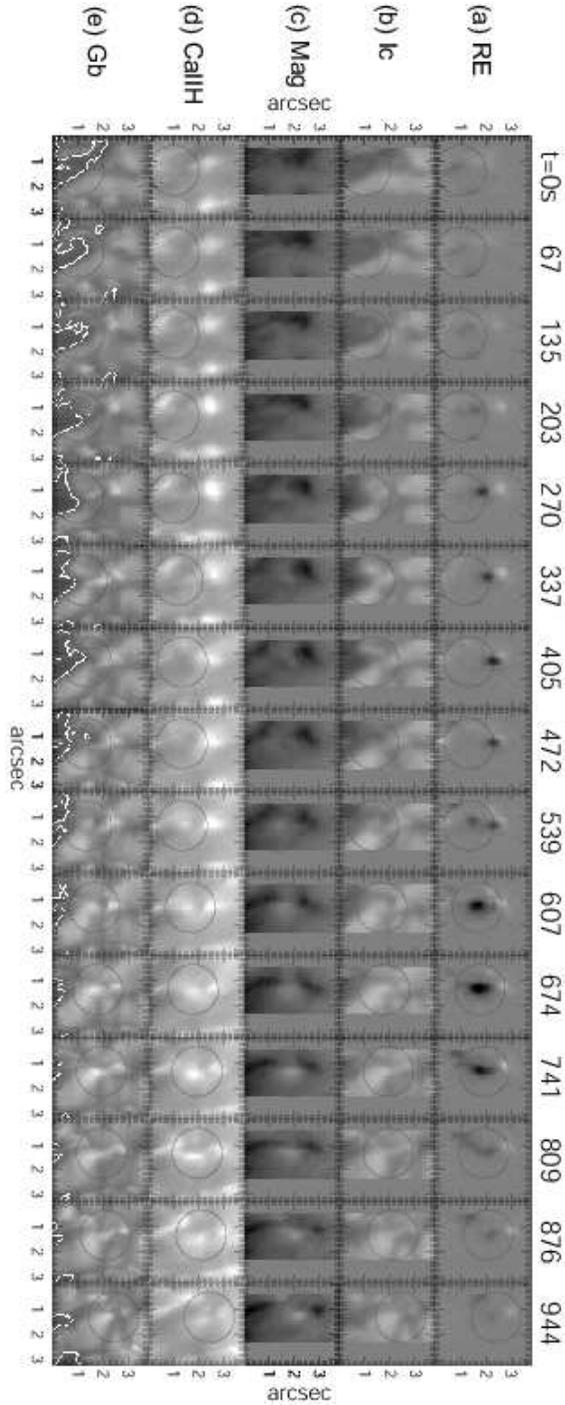}
\caption{Temporal evolution of one typical high-speed downflow event
observed at the sunspot penumbral boundary. (a) The measure for describing
signal excess at red wing, i.e., the integration of Stokes V signals
between 259 - 431 m\AA\ from the line center, (b) Stokes I continuum, (c)
magnetic flux, which is the wavelength integration of blue wing of Stokes
V, (d) Ca {\sc II H} and (e) G-band images from the BFI. The field of view is
3.8 (N-S) $\times$ 3.3 (E-W) arcsec though the quantities derived from
Stokes-polarimetric measurements (a, b, c) have a narrower field of view 
(2.1 arcsec). Circles with radius of 1.0 arcsec are given at each frame
and the Stokes profiles at the center position of the circles are shown
in Figure 5. The sunspot penumbral boundary is shown by a contour 
in the G-band images.}
\end{figure}

\begin{figure}
\epsscale{1}
\plotone{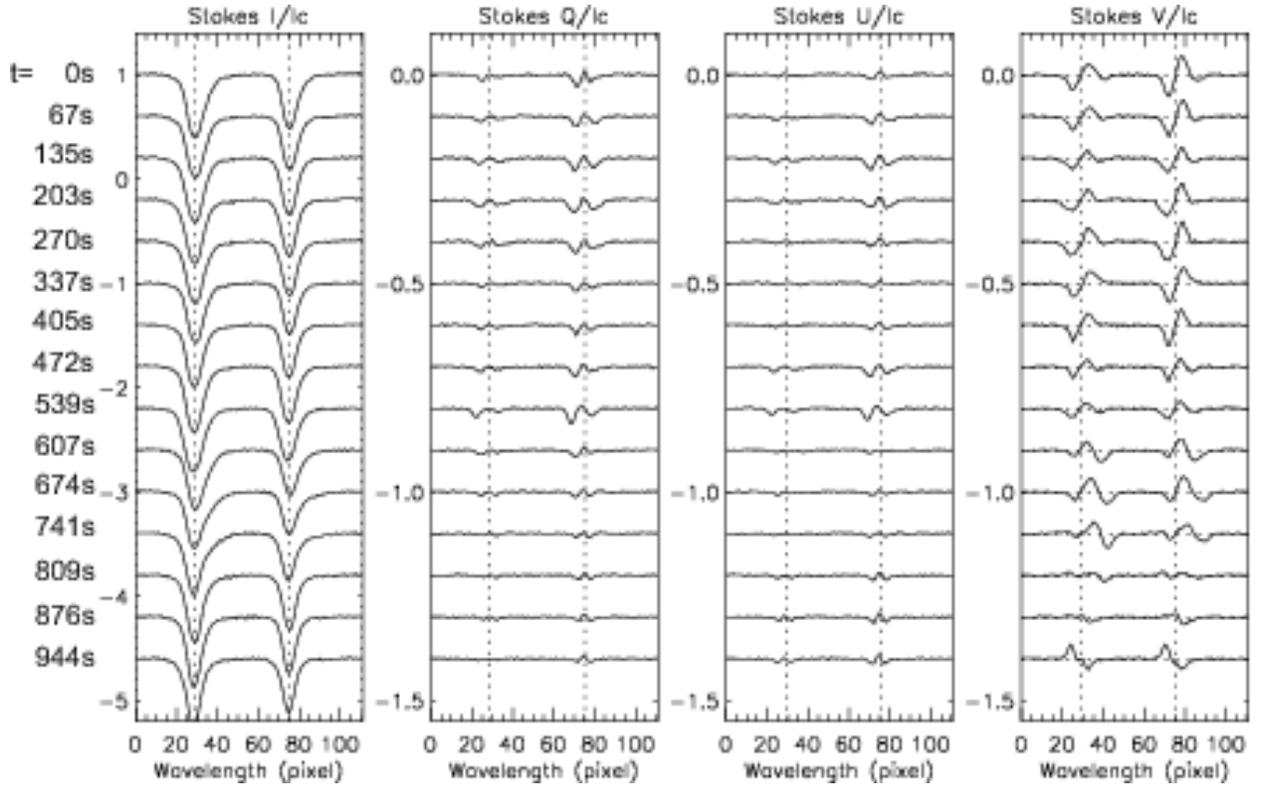}
\caption{Temporal evolution of the full Stokes profiles of Fe {\sc I} lines at
630.15 and 630.25 nm observed at the location of the high-speed downflow
event shown in Figure 4. The profiles are normalized to the nearby continuum.
The horizontal axis is wavelength position with spectral sampling of 
21.549 m\AA\ /pixel. Dotted lines shows the averaged wavelength position
of the spectral lines.}
\end{figure}

\begin{figure}
\epsscale{0.45}
\plotone{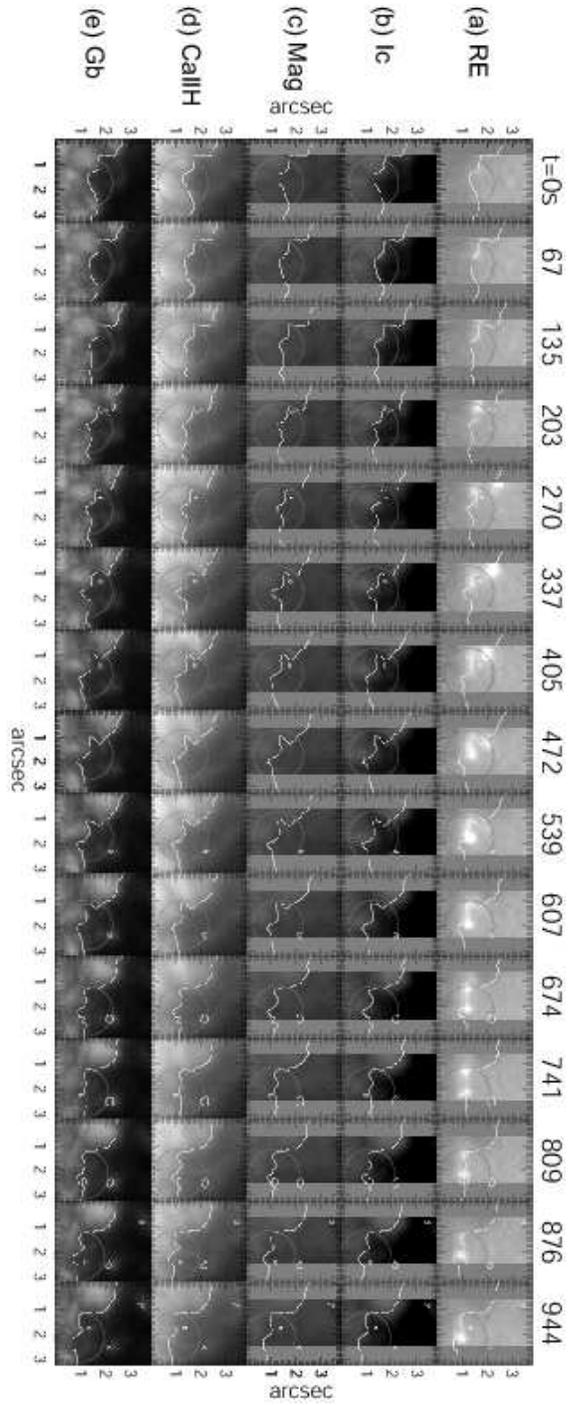}
\caption{Same as Figure 4 for one typical high-speed downflow event observed
  at the umbral boundary. The sunspot umbral boundary is shown by the G-band 
  intensity contours drawn in all the panels.}
\end{figure}

\begin{figure}
\epsscale{1}
\plotone{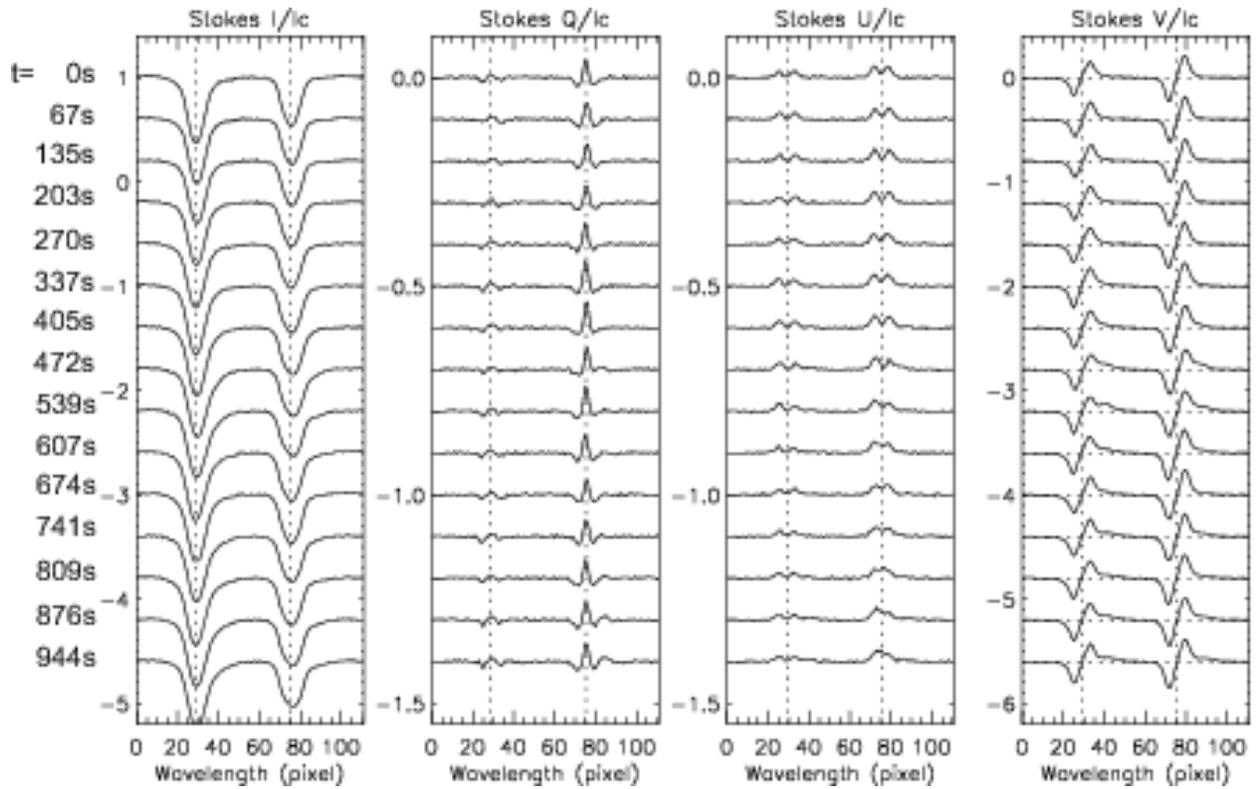}
\caption{Temporal evolution of the full Stokes profiles of Fe {\sc I} lines
at 630.15 and 630.25 nm observed at the location of the high-speed downflow
event shown in Figure 6. The vertical scale of Stokes V profiles is 4 times
larger than that of Stokes Q and U. Otherwise the same as Figure 5. }
\end{figure}

\begin{figure}
\epsscale{0.45}
\plotone{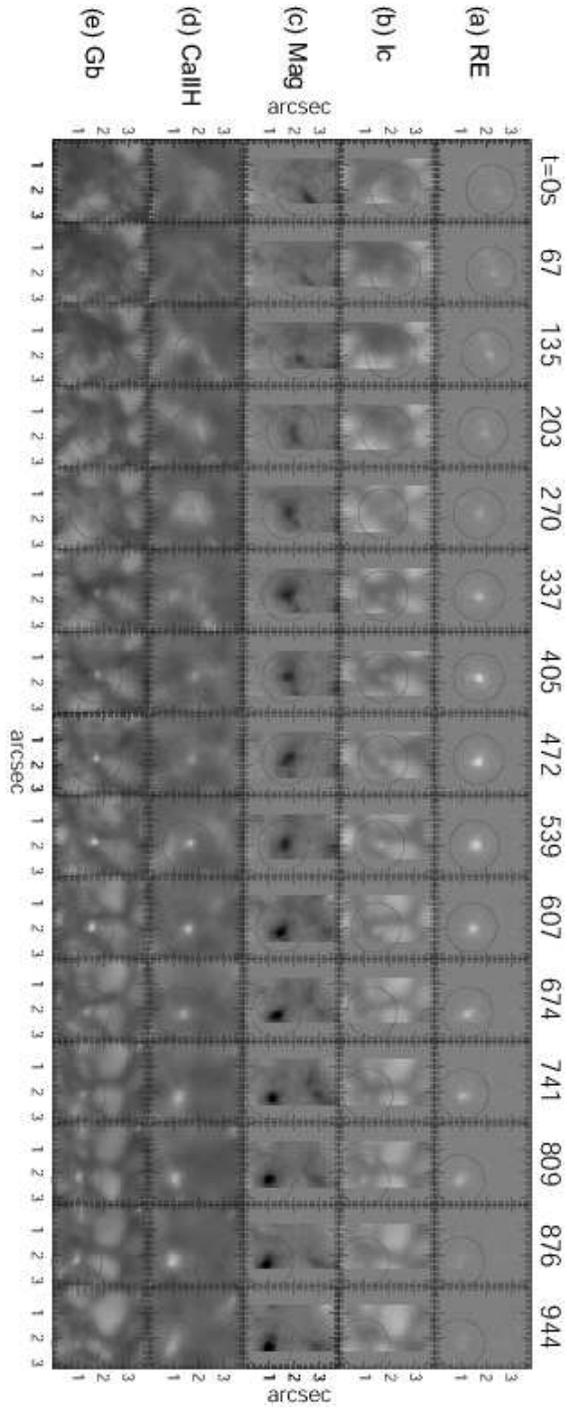}
\caption{Same as Figure 4 for one typical high-speed downflow event observed
in the moat region far from the sunspot penumbra.}
\end{figure}

\begin{figure}
\epsscale{1}
\plotone{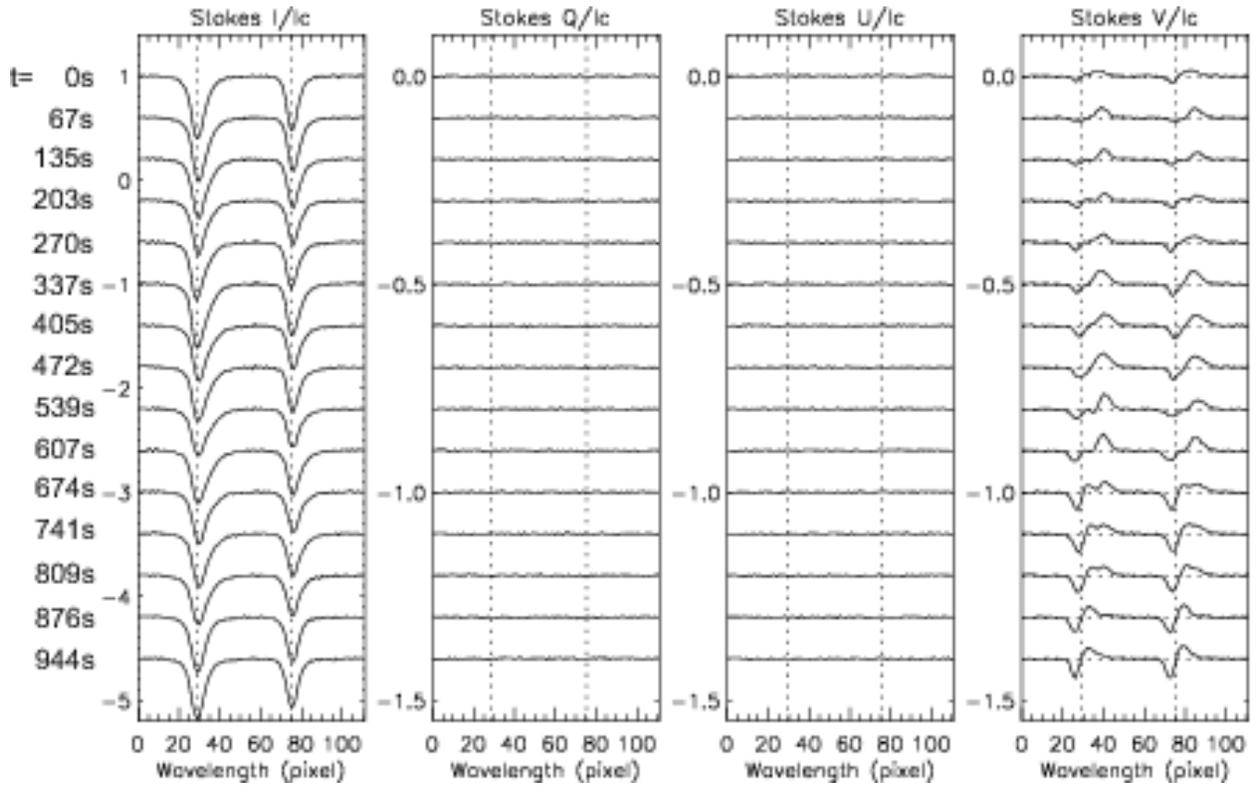}
\caption{Temporal evolution of the full Stokes profiles of Fe {\sc I} lines
observed at the location of the high-speed downflow
event shown in Figure 8. Otherwise the same as Figure 5. }
\end{figure}

\begin{figure}
\epsscale{1.0}
\plotone{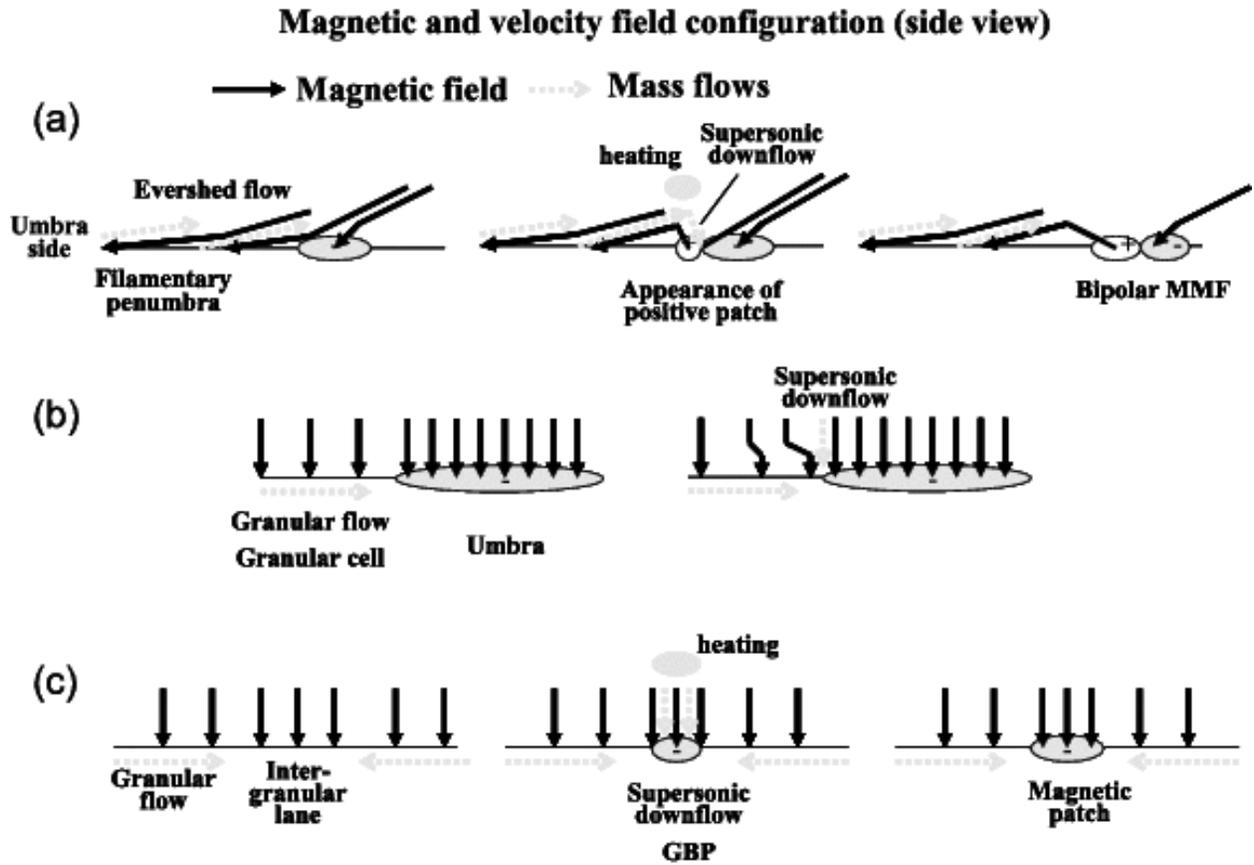}
\caption{Magnetic and velocity field configurations are schematically 
 shown to interpret the observed downflows at the three locations.}
\end{figure}

\end{document}